\documentclass[amsmath,notitlepage,amssymb,aps,showkeys,floatfix,prd,a4paper,
  twocolumn,nofootinbib]{revtex4-1}

\usepackage{graphicx}
\usepackage{amsmath}
\usepackage{epstopdf}
\usepackage{amsfonts,amssymb}
\usepackage{float}
\usepackage[utf8]{inputenc}
\usepackage{pstricks}
\usepackage{color}
\usepackage[compat=1.0.0]{tikz-feynman}
\usepackage{simplewick}
\usepackage{color, colortbl}
\definecolor{Gray}{gray}{0.9}
\usepackage{float}
\usepackage{tikz-feynman}
\usepackage{feynmp}
\usepackage{forest}

\usepackage{verbatim}    
\usepackage{dcolumn}
\usepackage{bm}
\usepackage{epsfig}
\usepackage{braket}
\usepackage[normalem]{ulem} 
\usepackage{longtable,booktabs}
\usepackage{booktabs}

\newcommand{\be}{\begin{equation}}
\newcommand{\ee}{\end{equation}}
\newcommand{\ben}{\begin{eqnarray}}
\newcommand{\een}{\end{eqnarray}}

\newcommand{\pslash}{\not{\hbox{\kern-2.3pt $p$}}}
\newcommand{\pdslash}{\not{\hbox{\kern-2pt $\partial$}}}

\begin{document}

\title{Hadronic scattering effects on $\psi(2S)$  suppression in relativistic heavy-ion collisions }

\author{ L. M. Abreu$^{1,2}$} \email{luciano.abreu@ufba.br}
\author{F. S. Navarra$^{2}$} \email{navarra@if.usp.br}
\author{H. P. L. Vieira$^{1}$} \email{hildeson.paulo@ufba.br}

\affiliation{$^{1}$Instituto de F\'isica, Universidade Federal da Bahia,
Campus Universit\'ario de Ondina, 40170-115, Bahia, Brazil}

\affiliation{$^{2}$Instituto de F\'{\i}sica, Universidade de S\~{a}o Paulo, Rua do Mat\~ao, 1371, CEP 05508-090,  S\~{a}o Paulo, SP, Brazil}


\begin{abstract}

In this work we estimate the $N_{\psi(2S)} / N_{J/\psi}$ yield ratio in heavy-ion collisions, considering the interactions of the $\psi (2S) $ and $J/\psi$ states with light mesons in the  hadron gas formed at the late stages of these collisions. Starting from the appropriate effective Lagrangians, we first compute the thermally-averaged cross sections for the production and absorption of the mentioned states, and then use 
them as input in the rate equations to determine the time evolution of  $N_{\psi(2S)}$, $N_{J/\psi}$ and $N_{\psi(2S)} / N_{J/\psi}$. The main
conclusion of our study is that the $\psi (2S) $ and $J/\psi$ multiplicities do not change much in the hadron gas phase and that 
the $\psi (2S)$ is more absorbed than the  $J/\psi$.  The obtained final ratio is in  qualitative agreement with experimental data.  

\end{abstract}

\date{\today}


\maketitle

\section{Introduction}
\label{Introduction} 

\bigskip

The study of charmonium states, such as the $J/\psi$ and $\psi(2S)$  remains a hot topic in high energy heavy ion collisions. 
Heavy quarkonium  production is strongly affected by the formation of  quark-gluon plasma (QGP), making the $Q \bar Q$ bound states relevant probes of this deconfined state of matter. The production of  $J/\psi$ has been largely investigated over the last decades. The relative suppression of $J/\psi$-production reported in nucleus-nucleus collisions with respect to the proton–proton collisions (scaled by the number of binary collisions) at SPS ~\cite{NA50:2004sgj}, at RHIC ~\cite{PHENIX:2006gsi,PHENIX:2011img,STAR:2009irl,STAR:2013eve} and at  LHC~\cite{CMS:2012bms,ALICE:2012jsl,ALICE:2013osk,ALICE:2015jrl,ALICE:2016flj,ALICE:2019lga,ALICE:2019nrq,ATLAS:2018hqe,CMS:2017uuv,
ALICE:2023gco}  was regarded as a sign of the dissociation in the QGP, caused by color Debye screening~\cite{Matsui:1986dk}. However, at LHC energies the $J/\psi$ suppression was found to be less pronounced than at RHIC energies (especially at low transverse momentum)~\cite{ALICE:2015jrl}, in contrast to the color screening predictions. Therefore, an additional $J/\psi$ regeneration mechanism  was  proposed to account for the formation of $J/\psi$ mesons via the coalescence of uncorrelated charm quarks and antiquarks~\cite{Thews:2000rj,rapp01,rapp22}. 

The  production of the $\psi(2S)$  state is of particular interest,  as the magnitude of medium effects may be much stronger due to its larger size and weaker binding, making its dissociation easier when compared to the $J/\psi$. Indeed, $\psi(2S)$ production in heavy ion collisions has been reported to be more suppressed than the  $J/\psi$ one~\cite{ALICE:2015jrl,CMS:2014vjg,Leoncino:2015ooa,CMS:2016wgo,MartinBlanco:2017njc,ALICE:2020vjy,PHENIX:2022nrm}).  One interesting observable in this analysis is the double ratio $ (N_{\psi(2S)}/N_{J/\psi})_{PbPb}/(N_{\psi(2S)}/N_{J/\psi})_{pp}$, i.e. the ratio of the corresponding nuclear modification factors, which allows for  the (at least partial) cancellation of corrections related to acceptance, efficiency, and integrated luminosity as well as their associated uncertainties. As discussed in Ref.~\cite{CMS:2016wgo}, this double ratio is affected by the competition of the Debye screening and regeneration effects: while the former could make the  double ratio smaller than unity, the latter  should make it exceed unity if uncorrelated quark coalescence produces more $\psi(2S)$'s  than $J/\psi$'s. 
The data from Ref.~\cite{CMS:2016wgo} have shown that in the measured bins the double ratio is lower than unity, suggesting that the $\psi(2S)$ multiplicity is more suppressed than that of the $J/\psi$. Very recently, in \cite{alice24} the  ratio  
$ N_{\psi(2S)} \, BR / N_{(J/\psi)} \, BR$  was measured by the ALICE Collaboration ($BR$ is the $J/\psi (\psi(2S))$  branching ratio into
$\mu^+ \mu^-$). This ratio was correctly reproduced by the TAMU transport model \cite{rapp24}. 
 
All the effects mentioned above refer to the QGP. Ideally,  measuring the $\psi(2S)$  and $J/\psi$ yields we may obtain information about the 
QGP. However, as the QGP expands, cools down and hadronizes, the early-formed quarkonia  must propagate through the medium and interact with 
the light mesons in the hadron gas (HG). These interactions could distort or even wash out all the valuable information coming from the QGP. 

Although quite reasonable, until recently the existence of a hadron gas phase was never experimentally proved. In \cite{alice-ks} the existence
of the hadron gas phase received strong support from the observation of $K^*$ suppression. This particle is very special because it has a lifetime of 
$~4$ fm/c, being shorter than the duration of the hadron gas phase, which is believed to be of the order of
$10$ fm/c. When the decay $ K^* \to K \, \pi$ happens in the hadronic medium, the daughter particles, $K$ and $\pi$, interact further with
other particles in the environment, changing their energy and momentum, and even if they can be measured at the
end of the heavy ion collision, the invariant mass of the pair is no longer equal to the $K^*$ mass. The $K^*$ which is no
longer reconstructed is lost and we  observe a reduction in the final yield of this resonance, which can 
be attributed to the existence of the hadron gas phase.  The fate of the $K^*$ in a hadronic medium was carefully studied in 
\cite{chiara21} where the ALICE data were very well  reproduced.  Taken together with previous measurements and previous theoretical estimates
along the same line, these works constitute strong evidence of the hadron gas formation.

In a hadron gas the overall multiplicities can suffer modifications due to  production and absorption processes, as was pointed out in previous works~\cite{xint-14-16,jpsi-18,dstar22,tcc22,zczcs,chi4274,psi2s}. In particular, in Ref.~\cite{jpsi-18} the  $J/\psi$ dissociation and production reactions with light mesons have been studied and the corresponding changes in the $J/\psi$  multiplicity have been evaluated. 
On the other hand, similar evaluations for the $\psi(2S)$ are scarce in the literature. Very recently, in Ref.~\cite{psi2s}, the $\psi(2S)$ interactions in a hadronic medium were investigated and the modifications on the  $X(3872)$ to $\psi(2S)$ yield ratio due to the dissociation/absorption by hadronic interactions
were studied. However, with the exception of  ~\cite{rapp24} no equivalent analysis has been developed for the $\psi(2S)$ to $J/\psi$ yield ratio. 

The purpose of this work is to estimate the ratio
\be
\mathcal{R}_{\psi} = \frac{N_{\psi(2S)}}{N_{J/\psi}}
\label{rat}
\ee 
in heavy-ion collisions, considering the interactions of the $\psi (2S) $ and $J/\psi$ states with light mesons in the hadron gas formed at the late stages of these collisions. To this end we update the
study of $J/\psi$ production performed in \cite{jpsi-18} and combine it with the analogous study of $\psi(2S)$ production concluded recently
\cite{psi2s}. In these works we use the thermally-averaged cross sections for the production and absorption of the mentioned states (obtained from effective Lagrangians) which serve as input in the rate equations to determine the $\psi(2S)$ and $J/\psi$ final yields and their ratio. We also evaluate the dependence of these quantities with the charged-particle pseudorapidity density at mid-rapidity, $\mathcal{N} \equiv [dN_{ch}/d\eta \,(|\eta|<0.5)]^{1/3}$, which can be related to the size of the system. 
The formalism employed will be briefly described in the next sections. All details can be found in the references.

\section{Charmonium cross section}
\label{Framework} 

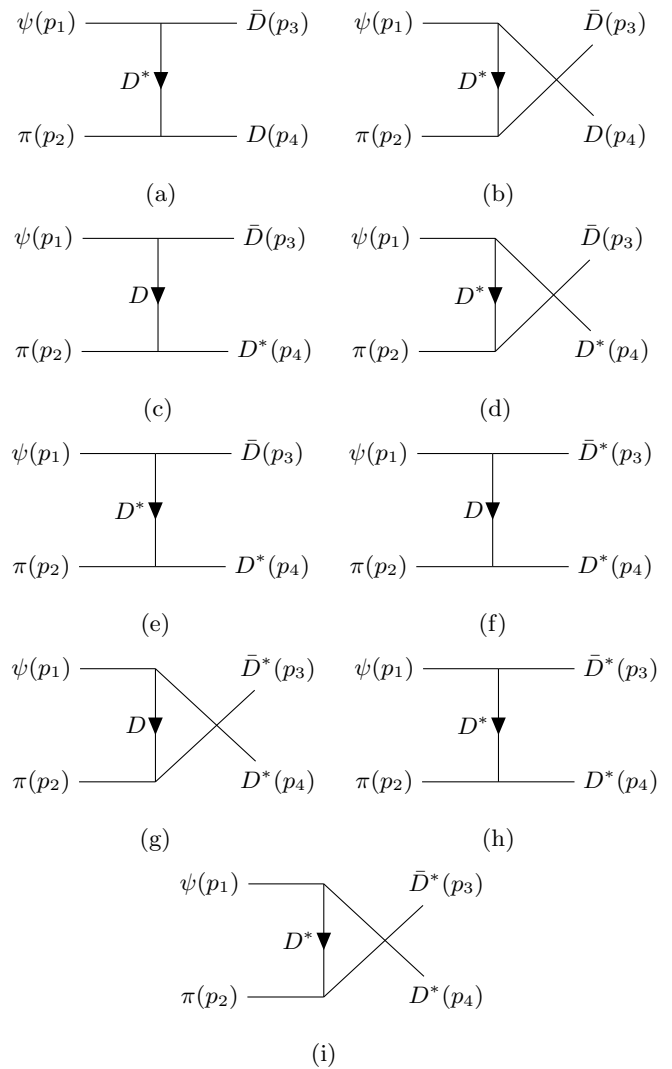
\begin{figure}[!htbp]
	\centering
\begin{tikzpicture}
\begin{feynman}
\vertex (a1) {$\psi (p_{1})$};
	\vertex[right=1.5cm of a1] (a2);
	\vertex[right=1.cm of a2] (a3) {$\bar{D} (p_{3})$};
	\vertex[right=1.4cm of a3] (a4) {$\psi (p_{1})$};
	\vertex[right=1.5cm of a4] (a5);
	\vertex[right=1.cm of a5] (a6) {$\bar{D} (p_{3})$};
\vertex[below=1.5cm of a1] (c1) {$\pi (p_{2})$};
\vertex[below=1.5cm of a2] (c2);
\vertex[below=1.5cm of a3] (c3) {$D (p_{4})$};
\vertex[below=1.5cm of a4] (c4) {$\pi (p_{2})$};
\vertex[below=1.5cm of a5] (c5);
\vertex[below=1.5cm of a6] (c6) {$D (p_{4})$};
	\vertex[below=2cm of a2] (d2) {(a)};
	\vertex[below=2cm of a5] (d5) {(b)};
\diagram* {
  (a1) -- (a2), (a2) -- (a3), (c1) -- (c2), (c2) -- (c3), (a2) --
  [fermion, edge label'= $D^{*}$] (c2), (a4) -- (a5), (a5) -- (c6),
  (c4) -- (c5), (c5) -- (a6), (a5) -- [fermion, edge label'= $D^{*}$] (c5)
}; 
\end{feynman}
\end{tikzpicture}

\begin{tikzpicture}
\begin{feynman}
\vertex (a1) {$\psi (p_{1})$};
	\vertex[right=1.5cm of a1] (a2);
	\vertex[right=1.cm of a2] (a3) {$\bar{D} (p_{3})$};
	\vertex[right=1.4cm of a3] (a4) {$\psi (p_{1})$};
	\vertex[right=1.5cm of a4] (a5);
	\vertex[right=1.cm of a5] (a6) {$\bar{D} (p_{3})$};
\vertex[below=1.5cm of a1] (c1) {$\pi (p_{2})$};
\vertex[below=1.5cm of a2] (c2);
\vertex[below=1.5cm of a3] (c3) {$D^{*} (p_{4})$};
\vertex[below=1.5cm of a4] (c4) {$\pi (p_{2})$};
\vertex[below=1.5cm of a5] (c5);
\vertex[below=1.5cm of a6] (c6) {$D^{*} (p_{4})$};
	\vertex[below=2cm of a2] (d2) {(c)};
	\vertex[below=2cm of a5] (d5) {(d)};
\diagram* {
  (a1) -- (a2), (a2) -- (a3), (c1) -- (c2), (c2) -- (c3), (a2) --
  [fermion, edge label'= $D$] (c2), (a4) -- (a5), (a5) -- (c6),
  (c4) -- (c5), (c5) -- (a6), (a5) -- [fermion, edge label'= $D^{*}$] (c5)
}; 
\end{feynman}
\end{tikzpicture}

\begin{tikzpicture}
\begin{feynman}
\vertex (a1) {$\psi (p_{1})$};
	\vertex[right=1.5cm of a1] (a2);
	\vertex[right=1.cm of a2] (a3) {$\bar{D} (p_{3})$};
	\vertex[right=1.4cm of a3] (a4) {$\psi (p_{1})$};
	\vertex[right=1.5cm of a4] (a5);
	\vertex[right=1.cm of a5] (a6) {$\bar{D}^{*} (p_{3})$};
\vertex[below=1.5cm of a1] (c1) {$\pi (p_{2})$};
\vertex[below=1.5cm of a2] (c2);
\vertex[below=1.5cm of a3] (c3) {$D^{*} (p_{4})$};
\vertex[below=1.5cm of a4] (c4) {$\pi (p_{2})$};
\vertex[below=1.5cm of a5] (c5);
\vertex[below=1.5cm of a6] (c6) {$D^{*} (p_{4})$};
	\vertex[below=2cm of a2] (d2) {(e)};
	\vertex[below=2cm of a5] (d5) {(f)};
\diagram* {
  (a1) -- (a2), (a2) -- (a3), (c1) -- (c2), (c2) -- (c3), (a2) --
  [fermion, edge label'= $D^{*}$] (c2), (a4) -- (a5), (a5) -- (a6),
  (c4) -- (c5), (c5) -- (c6), (a5) -- [fermion, edge label'= $D$] (c5)
}; 
\end{feynman}
\end{tikzpicture}

\begin{tikzpicture}
\begin{feynman}
\vertex (a1) {$\psi (p_{1})$};
	\vertex[right=1.5cm of a1] (a2);
	\vertex[right=1.cm of a2] (a3) {$\bar{D}^{*} (p_{3})$};
	\vertex[right=1.4cm of a3] (a4) {$\psi (p_{1})$};
	\vertex[right=1.5cm of a4] (a5);
	\vertex[right=1.cm of a5] (a6) {$\bar{D}^{*} (p_{3})$};
\vertex[below=1.5cm of a1] (c1) {$\pi (p_{2})$};
\vertex[below=1.5cm of a2] (c2);
\vertex[below=1.5cm of a3] (c3) {$D^{*} (p_{4})$};
\vertex[below=1.5cm of a4] (c4) {$\pi (p_{2})$};
\vertex[below=1.5cm of a5] (c5);
\vertex[below=1.5cm of a6] (c6) {$D^{*} (p_{4})$};
	\vertex[below=2cm of a2] (d2) {(g)};
	\vertex[below=2cm of a5] (d5) {(h)};
\diagram* {
  (a1) -- (a2), (a2) -- (c3), (c1) -- (c2), (c2) -- (a3), (a2) --
  [fermion, edge label'= $D$] (c2), (a4) -- (a5), (a5) -- (a6),
  (c4) -- (c5), (c5) -- (c6), (a5) -- [fermion, edge label'= $D^{*}$] (c5)
}; 
\end{feynman}
\end{tikzpicture}

\begin{tikzpicture}
\begin{feynman}
\vertex (a1) {$\psi (p_{1})$};
	\vertex[right=1.5cm of a1] (a2);
	\vertex[right=1.cm of a2] (a3) {$\bar{D}^{*} (p_{3})$};
\vertex[below=1.5cm of a1] (c1) {$\pi (p_{2})$};
\vertex[below=1.5cm of a2] (c2);
\vertex[below=1.5cm of a3] (c3) {$D^{*} (p_{4})$};
	\vertex[below=2cm of a2] (d2) {(i)};
\diagram* {
  (a1) -- (a2), (a2) -- (c3), (c1) -- (c2), (c2) -- (a3), (a2) --
  [fermion, edge label'= $D^{*}$] (c2)
}; 
\end{feynman}
\end{tikzpicture}
	\caption{Born diagrams for the processes   
$\psi \pi \rightarrow \bar{D} D$  [(a) and (b)],
$\psi \pi \rightarrow \bar{D}^{*} D$  [(c) - (e)],
and $\psi \pi \rightarrow \bar{D}^{*} D^{*}$  [(f) - (i)]. 
The particle charges are not specified. We denote $  \psi \equiv J/\psi, \psi(2S)  $. }
\label{DIAG1}
\end{figure}

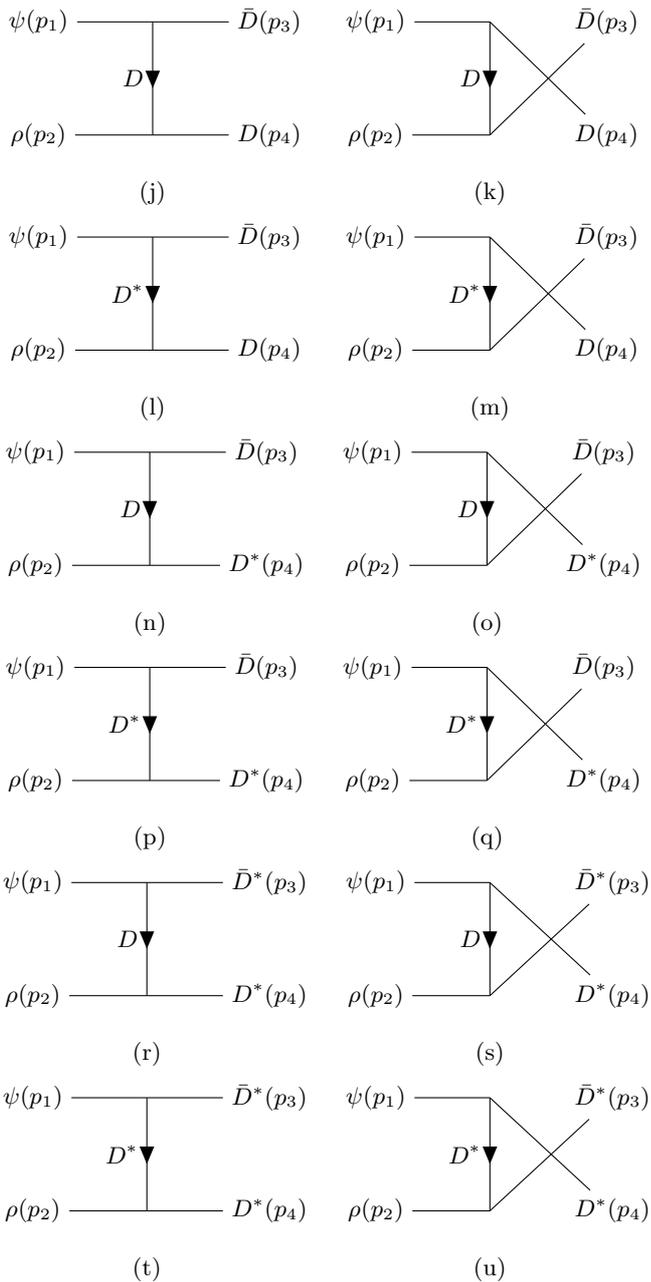
\begin{figure}[!htbp]
	\centering
\begin{tikzpicture}
\begin{feynman}
\vertex (a1) {$\psi (p_{1})$};
	\vertex[right=1.5cm of a1] (a2);
	\vertex[right=1.cm of a2] (a3) {$\bar{D} (p_{3})$};
	\vertex[right=1.4cm of a3] (a4) {$\psi (p_{1})$};
	\vertex[right=1.5cm of a4] (a5);
	\vertex[right=1.cm of a5] (a6) {$\bar{D} (p_{3})$};
\vertex[below=1.5cm of a1] (c1) {$\rho (p_{2})$};
\vertex[below=1.5cm of a2] (c2);
\vertex[below=1.5cm of a3] (c3) {$D (p_{4})$};
\vertex[below=1.5cm of a4] (c4) {$\rho (p_{2})$};
\vertex[below=1.5cm of a5] (c5);
\vertex[below=1.5cm of a6] (c6) {$D (p_{4})$};
	\vertex[below=2cm of a2] (d2) {(j)};
	\vertex[below=2cm of a5] (d5) {(k)};
\diagram* {
  (a1) -- (a2), (a2) -- (a3), (c1) -- (c2), (c2) -- (c3), (a2) --
  [fermion, edge label'= $D$] (c2), (a4) -- (a5), (a5) -- (c6),
  (c4) -- (c5), (c5) -- (a6), (a5) -- [fermion, edge label'= $D$] (c5)
}; 
\end{feynman}
\end{tikzpicture}

\begin{tikzpicture}
\begin{feynman}
\vertex (a1) {$\psi (p_{1})$};
	\vertex[right=1.5cm of a1] (a2);
	\vertex[right=1.cm of a2] (a3) {$\bar{D} (p_{3})$};
	\vertex[right=1.4cm of a3] (a4) {$\psi (p_{1})$};
	\vertex[right=1.5cm of a4] (a5);
	\vertex[right=1.cm of a5] (a6) {$\bar{D} (p_{3})$};
\vertex[below=1.5cm of a1] (c1) {$\rho (p_{2})$};
\vertex[below=1.5cm of a2] (c2);
\vertex[below=1.5cm of a3] (c3) {$D (p_{4})$};
\vertex[below=1.5cm of a4] (c4) {$\rho (p_{2})$};
\vertex[below=1.5cm of a5] (c5);
\vertex[below=1.5cm of a6] (c6) {$D (p_{4})$};
	\vertex[below=2cm of a2] (d2) {(l)};
	\vertex[below=2cm of a5] (d5) {(m)};
\diagram* {
  (a1) -- (a2), (a2) -- (a3), (c1) -- (c2), (c2) -- (c3), (a2) --
  [fermion, edge label'= $D^*$] (c2), (a4) -- (a5), (a5) -- (c6),
  (c4) -- (c5), (c5) -- (a6), (a5) -- [fermion, edge label'= $D^*$] (c5)
}; 
\end{feynman}
\end{tikzpicture}

\begin{tikzpicture}
\begin{feynman}
\vertex (a1) {$\psi (p_{1})$};
	\vertex[right=1.5cm of a1] (a2);
	\vertex[right=1.cm of a2] (a3) {$\bar{D} (p_{3})$};
	\vertex[right=1.4cm of a3] (a4) {$\psi (p_{1})$};
	\vertex[right=1.5cm of a4] (a5);
	\vertex[right=1.cm of a5] (a6) {$\bar{D} (p_{3})$};
\vertex[below=1.5cm of a1] (c1) {$\rho (p_{2})$};
\vertex[below=1.5cm of a2] (c2);
\vertex[below=1.5cm of a3] (c3) {$D^* (p_{4})$};
\vertex[below=1.5cm of a4] (c4) {$\rho (p_{2})$};
\vertex[below=1.5cm of a5] (c5);
\vertex[below=1.5cm of a6] (c6) {$D^* (p_{4})$};
	\vertex[below=2cm of a2] (d2) {(n)};
	\vertex[below=2cm of a5] (d5) {(o)};
\diagram* {
  (a1) -- (a2), (a2) -- (a3), (c1) -- (c2), (c2) -- (c3), (a2) --
  [fermion, edge label'= $D$] (c2), (a4) -- (a5), (a5) -- (c6),
  (c4) -- (c5), (c5) -- (a6), (a5) -- [fermion, edge label'= $D$] (c5)
}; 
\end{feynman}
\end{tikzpicture}

\begin{tikzpicture}
\begin{feynman}
\vertex (a1) {$\psi (p_{1})$};
	\vertex[right=1.5cm of a1] (a2);
	\vertex[right=1.cm of a2] (a3) {$\bar{D} (p_{3})$};
	\vertex[right=1.4cm of a3] (a4) {$\psi (p_{1})$};
	\vertex[right=1.5cm of a4] (a5);
	\vertex[right=1.cm of a5] (a6) {$\bar{D} (p_{3})$};
\vertex[below=1.5cm of a1] (c1) {$\rho (p_{2})$};
\vertex[below=1.5cm of a2] (c2);
\vertex[below=1.5cm of a3] (c3) {$D^* (p_{4})$};
\vertex[below=1.5cm of a4] (c4) {$\rho (p_{2})$};
\vertex[below=1.5cm of a5] (c5);
\vertex[below=1.5cm of a6] (c6) {$D^* (p_{4})$};
	\vertex[below=2cm of a2] (d2) {(p)};
	\vertex[below=2cm of a5] (d5) {(q)};
\diagram* {
  (a1) -- (a2), (a2) -- (a3), (c1) -- (c2), (c2) -- (c3), (a2) --
  [fermion, edge label'= $D^*$] (c2), (a4) -- (a5), (a5) -- (c6),
  (c4) -- (c5), (c5) -- (a6), (a5) -- [fermion, edge label'= $D^*$] (c5)
}; 
\end{feynman}
\end{tikzpicture}

\begin{tikzpicture}
\begin{feynman}
\vertex (a1) {$\psi (p_{1})$};
	\vertex[right=1.5cm of a1] (a2);
	\vertex[right=1.cm of a2] (a3) {$\bar{D}^* (p_{3})$};
	\vertex[right=1.4cm of a3] (a4) {$\psi (p_{1})$};
	\vertex[right=1.5cm of a4] (a5);
	\vertex[right=1.cm of a5] (a6) {$\bar{D}^* (p_{3})$};
\vertex[below=1.5cm of a1] (c1) {$\rho (p_{2})$};
\vertex[below=1.5cm of a2] (c2);
\vertex[below=1.5cm of a3] (c3) {$D^* (p_{4})$};
\vertex[below=1.5cm of a4] (c4) {$\rho (p_{2})$};
\vertex[below=1.5cm of a5] (c5);
\vertex[below=1.5cm of a6] (c6) {$D^* (p_{4})$};
	\vertex[below=2cm of a2] (d2) {(r)};
	\vertex[below=2cm of a5] (d5) {(s)};
\diagram* {
  (a1) -- (a2), (a2) -- (a3), (c1) -- (c2), (c2) -- (c3), (a2) --
  [fermion, edge label'= $D$] (c2), (a4) -- (a5), (a5) -- (c6),
  (c4) -- (c5), (c5) -- (a6), (a5) -- [fermion, edge label'= $D$] (c5)
}; 
\end{feynman}
\end{tikzpicture}

\begin{tikzpicture}
\begin{feynman}
\vertex (a1) {$\psi (p_{1})$};
	\vertex[right=1.5cm of a1] (a2);
	\vertex[right=1.cm of a2] (a3) {$\bar{D}^* (p_{3})$};
	\vertex[right=1.4cm of a3] (a4) {$\psi (p_{1})$};
	\vertex[right=1.5cm of a4] (a5);
	\vertex[right=1.cm of a5] (a6) {$\bar{D}^* (p_{3})$};
\vertex[below=1.5cm of a1] (c1) {$\rho (p_{2})$};
\vertex[below=1.5cm of a2] (c2);
\vertex[below=1.5cm of a3] (c3) {$D^* (p_{4})$};
\vertex[below=1.5cm of a4] (c4) {$\rho (p_{2})$};
\vertex[below=1.5cm of a5] (c5);
\vertex[below=1.5cm of a6] (c6) {$D^* (p_{4})$};
	\vertex[below=2cm of a2] (d2) {(t)};
	\vertex[below=2cm of a5] (d5) {(u)};
\diagram* {
  (a1) -- (a2), (a2) -- (a3), (c1) -- (c2), (c2) -- (c3), (a2) --
  [fermion, edge label'= $D^*$] (c2), (a4) -- (a5), (a5) -- (c6),
  (c4) -- (c5), (c5) -- (a6), (a5) -- [fermion, edge label'= $D^*$] (c5)
}; 
\end{feynman}
\end{tikzpicture}

	\caption{Born diagrams for the processes   
$\psi \rho \rightarrow \bar{D} D$  [(j) - (m)],
$\psi \rho \rightarrow \bar{D}^{*} D$  [(n) - (q)] and
$\psi \rho \rightarrow \bar{D}^{*} D^{*}$  [(r) - (u)].  The particle charges are not specified. 
We denote $ \psi \equiv J/\psi, \psi(2S) $.}
\label{DIAG2}
\end{figure}

We consider a hadronic medium constituted by the lightest and most abundant pseudoscalar and vector mesons, i.e. the pions and $\rho$ mesons.
The interaction between the charmonium states and these mesons will be described by effective Lagrangians. In the case of $J/\psi $  
interactions, this approach has a 25 years long history \cite{mu98,ko00,lee01,nava05,jpsi-18,mu23},  in which, curiously, the first 
\cite{mu98} and the latest work \cite{mu23}  were written  by the same author. From these previous works  we can conclude that the interactions with pions  give the dominant contribution. In the past, the interactions of the $\psi(2S)$ were never treated with effective Lagrangians. This was done for the first time in  Ref.~\cite{psi2s}, where
we made use of the similarities between the interactions involving the $\psi(2S)$ and those of the $J/\psi$. Applying the results found in 
\cite{jpsi-18} and in  \cite{psi2s} we study the reactions represented by the lowest-order Born diagrams contributing to the processes 
$\psi \pi \rightarrow \bar{D}^{(*)} D^{(*)}$ and $\psi \rho \rightarrow \bar{D}^{(*)} D^{(*)}$, as well as the inverse processes, where 
$\psi $ denotes the $J/ \psi$ or $\psi (2S)$ state. They are shown in Figs.~\ref{DIAG1} and~\ref{DIAG2}.

To calculate the relevant amplitudes, we make use of the  effective Lagrangians introduced in~\cite{ lee01, nava05, jpsi-18}, and summarized in~\cite{psi2s}. The form factors, employed to take into account the finite size of hadrons and prevent the artificial growth of the amplitudes with energy, have been calculated with QCD sum rules in Refs.~\cite{QCDSR:23,QCDSR:18, QCDSR:20, nava05}, and are displayed in Table I of~\cite{psi2s}. For the sake of conciseness, the effective Lagrangians and form factors will be not reproduced here. 
In the case of  the $\psi(2S)$, we adopt the following approach already presented in Ref.~\cite{psi2s}: we notice that in general the cross sections should be related to the size of the interacting hadron, and  remember that the first-excited charmonium state is larger than the fundamental one by a factor of about 2 (see the discussion in~\cite{Godfrey:1985xj}). From this geometrical argument we expect that the $\psi(2S)$ should have bigger cross sections than the $J/\psi$ by a factor of about 4. Hence, we use for the $\psi(2S)$ the same form factors obtained for the  $J/\psi$ but with the corresponding coupling constants taken in the range $[g_{J/\psi M_2 M_3}, 2g_{J/\psi M_2 M_3}]$ (with $M_2, M_3$ denoting the other particles in the vertex). This is the main source of uncertainty in the calculations.

The amplitudes corresponding to the diagrams shown in Figs.~\ref{DIAG1} and~\ref{DIAG2}  will be employed in the calculation of the respective vacuum cross section, which is defined in the center-of-mass (CM) frame as
\begin{eqnarray}
\sigma_{ab \rightarrow cd} (s) = \frac{1}{64 \pi^2 s g_{a} g_{b}} \frac{ | \vec{p}_{cd} |}{| \vec{p}_{ab} |}  \int d  \Omega \sum_{S, I} | \mathcal{M}_{ab \rightarrow cd} (s,\theta) |^2  ,
\label{eq:CrossSection}
\end{eqnarray} 
where $ ab \rightarrow cd $ denotes the process;  $\sqrt{s}$ is the CM energy; $|\vec{p}_{ab}|$ and $|\vec{p}_{cd}|$ are the magnitudes of three-momenta of initial and final particles in the CM frame, respectively; $\sum_{S,I} $ means the sum over the
spins and isospins of the particles in the initial and 
final state, weighted by the isospin and spin degeneracy factors $g_a = (2S_a + 1)(2I_a +  1)$ and  $g_b = (2S_b + 1)(2I_b +  1)$ of the two particles forming the initial state ~\cite{Workman:2022ynf}. 
For a discussion concerning the uncertainties in the cross sections we refer the reader to Ref.~\cite{jpsi-18} for the $J/\psi$ and  to Ref.~\cite{psi2s} for the $\psi(2S)$.

Since we are interested in reactions occurring in a hadron gas, the vacuum cross sections must be weighted by the thermal momentum 
distributions of the colliding particles. For the processes discussed above, the thermally-averaged cross section is given by~\cite{Koch:1986,lee13,exhic17}
\begin{eqnarray}
\langle \sigma_{a b \rightarrow c d } \upsilon_{a b}\rangle &  = & 
\frac{ \int  d^{3} \mathbf{p}_a  d^{3}
\mathbf{p}_b f_a(\mathbf{p}_a) f_b(\mathbf{p}_b) \sigma_{a b \rightarrow c d } 
\,\,\upsilon_{a b} }{ \int d^{3} \mathbf{p}_a  
d^{3} \mathbf{p}_b f_a(\mathbf{p}_a) f_b(\mathbf{p}_b) }
\nonumber \\
& = & \frac{1}{4 \beta_a ^2 K_{2}(\beta_a) \beta_b ^2 K_{2}(\beta_b) } 
\nonumber \\
& & \times \int _{z_0} ^{\infty } dz  K_{1}(z) \,\,\sigma_{ab \rightarrow cd}  (s=z^2 T^2) 
\nonumber \\
& & \times \left[ z^2 - (\beta_a + \beta_b)^2 \right]
\left[ z^2 - (\beta_a - \beta_b)^2 \right],
\nonumber \\
  \label{Eq:AvCrossSection}
\end{eqnarray}
where $\upsilon_{a b}$ is the relative velocity of the two initial interacting particles; the function $f_i(\mathbf{p}_i)$ is the Bose-Einstein distribution;
$\beta _i = m_i / T$, with $T$ being the temperature; $z_0 = max(\beta_a + \beta_b,\beta_c + \beta_d)$, and $K_1$ and $K_2$ are the modified Bessel functions of
second kind.  The average over the thermal momentum distributions washes out threshold   effects. 

\begin{figure}[!htbp]
	\centering
\includegraphics[{width=8.0cm}]{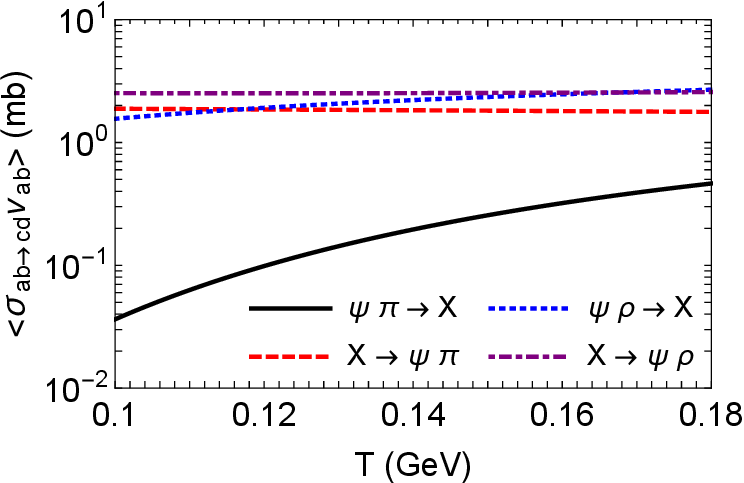} 
	\caption{Total thermally averaged cross sections for the absorption and production processes involving the channels $ J/ \psi \, \pi $ 
 and $J/ \psi \, \rho$, as a function of the temperature. $X$ stands for the sum of the open charm states $D \bar{D}$, $D^* \bar{D}$ and 
 $D^* \bar{D^*}$. }
 \label{fig3}
\end{figure}

\begin{figure}[!htbp]
	\centering
\includegraphics[{width=8.0cm}]{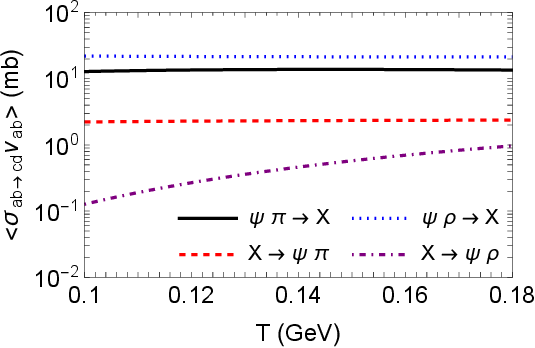} 
	\caption{Total thermally averaged cross sections for the absorption and production processes involving the channels 
 $ \psi(2S) \, \pi $ and $\psi(2S) \,  \rho$, as a function of the temperature. $X$ stands for the sum of the open charm states 
 $D \bar{D}$, $D^* \bar{D}$ and 
 $D^* \bar{D^*}$. }
\label{fig4}
\end{figure}


In Fig.~\ref{fig3}  we show the thermally-averaged cross sections as a function of the temperature for the $ J/ \psi$ absorption by light mesons and their inverse reactions using the cross-sections obtained previously. In Fig.~\ref{fig4} we show the analogous cross sections for 
$\psi(2S)$. They are taken from Refs.~\cite{jpsi-18} and \cite{psi2s}. For the sake of clarity we omit the theoretical uncertainty bands, which 
can be found in the mentioned works. 

From the figures we can draw a few important conclusions:

\noindent 
i) All the most important processes have nearly constant cross sections. All the strong variations close to the energy thresholds found in 
Refs. \cite{jpsi-18} and in \cite{psi2s} are washed out.  

\noindent
ii) The cross sections of the $J/\psi \, \pi$ processes are close to previous estimates made with QCD sum rules \cite{sigpi} and also  
with other effective theories \cite{mu98,ko00,lee01,nava05,mu23}. 

\noindent
iii) The $J/\psi$  is predominantly produced  while the  $\psi(2S)$ is predominantly absorbed.  This behavior is the dynamical reason for the 
decrease of the $N_{\psi(2S)}/N_{J/\psi}$ ratio.  If the system were in a box with constant volume and constant temperature, we would be able to read the final ratio from the above figures. However, the number of light and open charm mesons depends on the volume and on the temperature and the system is expanding and cooling. These effects will be taken into account in what follows.

\section{Rate equations}
\label{abundance}

In order to calculate  the multiplicities of $\psi (2S)$ and $J/\psi$ during the hadronic stage of heavy ion collisions we use the 
formalism  previously introduced in ~\cite{Koch:1986,lee13}. The cross sections 
derived in the previous section are used in the rate equations  
\begin{eqnarray} 
	\nonumber  \frac{ d N_{\psi}  (\tau)}{d \tau} &=& \sum_{\substack{\bar{c} = \bar{D},\bar{D}^{*}; \\ c = D, D^{*}; \\ 
\varphi = \pi, \rho}} 
\left[ \langle \sigma_{\bar{c}  c \rightarrow \varphi \psi} \upsilon_{\bar{c}  c} \rangle n_{\bar{c}  } (\tau) N_{c} (\tau)
 \right. 
	\nonumber \\ 
	&& -  \left.  \langle \sigma_{ \psi \varphi \rightarrow \bar{c}  c} \upsilon_{\psi \varphi } \rangle n_{\varphi} (\tau) N_{\psi}(\tau) 
\right], 
\label{Eq:RatioEquation}
\end{eqnarray}
where $N_{\psi}(\tau )$ denotes the multiplicity of the state $\psi$ $(\psi = \psi (2S), J/\psi )$; $n_i (\tau )$ and $N_i (\tau )$ 
represent the density and the number of mesons of type $i$ at a given $\tau $. 
The production and absorption thermally-averaged cross sections  are used as entries into the gain and loss terms, written in the first and second lines of Eq.~(\ref{Eq:RatioEquation}) respectively. 
We assume that all the particles involved in the reactions are in thermal equilibrium, and that the density $n_{i} (\tau)$ is written in the Maxwell-Boltzmann approximation as
\begin{eqnarray}
n_{i} (\tau) & =  & \frac{1}{2 \pi^2} \, \gamma_{i} \,  g_{i} \,  m_{i}^2 \, T(\tau) \, K_{2} \left(\frac{m_{i} }{T(\tau)}\right), 
\label{Eq:Densities}
\end{eqnarray}
where $\gamma _i$, $g_i$ and $m_i$ are the fugacity, degeneracy factor and mass of the particle $i$, respectively. The multiplicity $N_i (\tau)$ is then obtained by multiplying $n_i(\tau)$ by the volume.  

The hadron gas expansion is modeled by the boost invariant Bjorken flow with an accelerated transverse expansion, in which the volume $ V(\tau) $ and temperature $ T(\tau) $ profiles become functions of the proper time  $\tau$. Their parametrizations are fixed in order to describe a hadronic medium formed in central PbPb collisions at CM energy $\sqrt{s_{NN}} = 5.02$ TeV. More details can be found in ~\cite{psi2s,exhic17}. 
Accordingly, the multiplicities of the light mesons as well as the multiplicity of charm quarks (which is considered approximately constant during the hydrodynamic expansion) in charmed mesons are also shown in Table~II of~\cite{psi2s}.

\section{Initial conditions} 

The initial condition for the multiplicity $N_{\psi}  (\tau)$ in the kinetic equation (\ref{Eq:RatioEquation}) is the multiplicity of a state $\psi$ at the end of QGP. It is  very difficult to estimate this quantity, given the non-perturbative nature of the hadronization process. Two 
very different models are used for this purpose: the coalescence model (CM) \cite{exhic17} and the Statistical Hadronization Model (SHM)
\cite{shm-19}. In these two models
it is assumed that no bound states survive in the plasma and that all the charmonium states are formed during hadronization. In the CM the 
abundance of charmonium state is related to the overlap between its wave function and the phase space distribution of quarks and anti-quarks
at the phase transition. In this model,  the wave functions and quantum numbers play a crucial role. In the SHM, the multiplicities are essentially given by Eq.~(\ref{Eq:Densities}), in which the relevant quantities are the mass, degeneracy factor, temperature and chemical potential.   Moreover in the SHM it is assumed that the produced particles are in chemical equilibrium. This means that even if there is a long hadron gas phase, the multiplicities will be conserved because there is a  (assumed) compensation between production and absorption. In fact, our previous  calculations show that in some cases, such as in Ref.~\cite{dstar22}, the particle abundancies do not change in the hadron gas phase, i.e. they are already in chemical equilibrium soon after hadronization. 

As an alternative to the "fresh-start-at-hadronization"  scenario admitted in the CM and in the SHM, 
it is also possible that, after being destroyed, charmonium  states are formed again, or "regenerated", in the QGP phase. As a consequence, 
at hadronization, the multiplicities will be obtained from the solution of a rate equation with gain and loss terms \cite{rapp24},  in the 
same way as we do in this work for the hadron gas phase.

As initial conditions for our analysis, we will assume that the multiplicities 
are given by the SHM.  Then, for the central values we find
\begin{eqnarray}
\nonumber N_{\psi(2S)} (\tau_H) & = &   7\times 10^{-3}, \\
 N_{J/\psi} (\tau_H) & =  & 0.25. 
\label{coalmod2}
\end{eqnarray}
It should be also mentioned that the value obtained for $ N_{\psi(2S)} (\tau_H)$ in Eq.~(\ref{coalmod2}) is higher than the one reported in Ref.~\cite{psi2s}, where it was computed with the CM  and found to be $ N_{\psi(2S)} (\tau_H)  =  1.8 \times 10^{-4} $.  
%
%
The value obtained for $ N_{J/\psi} (\tau_H)$ in Eq.~(\ref{coalmod2}) is smaller than the one found in 
Ref.~\cite{jpsi-18} ($N_{J/\psi} (\tau_H) = 1.67 $), where it was calculated with the SHM using a constant charm-quark-fugacity.
Here the total number of charm quarks in charmed hadrons is assumed to be approximately conserved during the hydrodynamic expansion, i.e. $ N_c = n_c (\tau) \times V(\tau) = const$ ~implying  a time-dependent charm-quark-fugacity factor $\gamma _c \equiv \gamma _c (\tau)$ appearing in the density $n_{i} (\tau)$ in Eq.~(\ref{Eq:Densities}).

\begin{figure}[!htbp]
	\centering
\includegraphics[{width=8.0cm}]{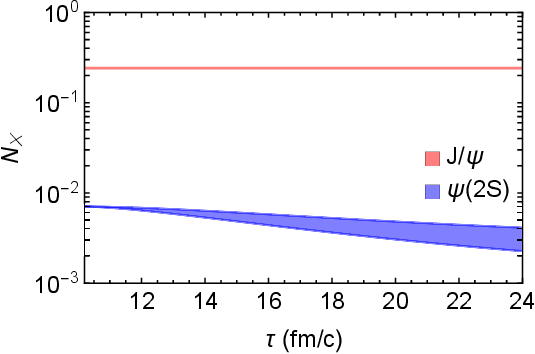} \\
\includegraphics[{width=8.0cm}]{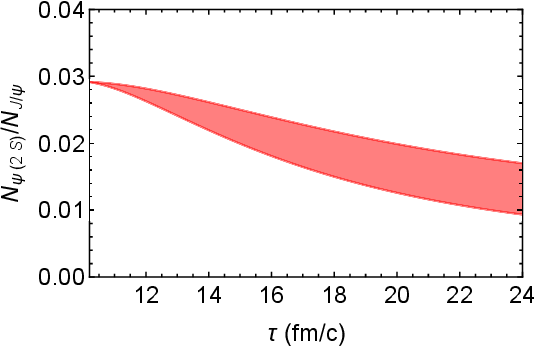} 
\caption{Multiplicity of $\psi$ and $\psi (2S)$ (upper panel) and their ratio (lower panel) as a function of the proper time in central PbPb collisions at $\sqrt{s_{NN}} =$ 5.02 TeV.  The bands show the uncertainties associated to changes in the  $\psi(2S)$ coupling constants in the range $[g_{J/\psi M_2 M_3}, 2g_{J/\psi M_2 M_3}]$.}  
\label{fig5}
\end{figure}
\section{Time evolution}

In Fig.~\ref{fig5} we show the time evolution of the $\psi(2S)$ and $J/\psi$ multiplicities as a function of the proper time. The final yield of the $\psi(2S)$ decreases by a factor of about two, revealing the dominant role of the loss terms at higher temperatures in the evolution equation (\ref{Eq:RatioEquation}). As the time evolves, the gain and loss terms become  of the same order (considering the uncertainties), and the multiplicities tend to stabilize, reaching chemical equilibrium.  The behavior seen in Fig.~\ref{fig5} is a consequence of  
the dominance of the  $\psi(2S)$ absorption over the production cross section, as seen in Fig.~\ref{fig4}.  In contrast, as seen in 
Fig.~\ref{fig3}, the $J/\psi$ production cross section is larger than the absorption one and hence its multiplicity remains approximately 
constant.  The bands in Fig.~\ref{fig5} show what happens to the particle abundances when we vary the 
$\psi(2S)$ coupling constants in the range $[g_{J/\psi M_2 M_3}, 2g_{J/\psi M_2 M_3}]$.

What will happen if we change the initial abundances by one order of magnitude or more ?  If we decrease them, we also decrease the loss 
term  in Eq.~(\ref{Eq:RatioEquation}). As a consequence,  at the beginning of the evolution we will see a growth of the abundance.   In  \cite{psi2s}, with a much smaller initial abundance, the growth was very pronounced.  However, in the end of the hadron gas phase 
the $\psi(2S)$ final yield plotted in Fig.~\ref{fig5}  and that reported in \cite{psi2s}  are approximately equal, considering the uncertainties. Therefore, we can conclude that in these Pb-Pb collisions the hadron gas phase lives long enough for the $\psi(2S)$ to 
reach approximate chemical equilibrium, loosing the memory of the initial conditions.  Of course this conclusion relies on the interaction 
cross sections which were employed. 

We also notice that, when compared to the findings reported in~\cite{jpsi-18} for central Pb-Pb collisions at $\sqrt{s_{NN}} =$ 5 TeV, the $J/\psi$ final yield in the present work is smaller. This comes from  the different values of the initial conditions already discussed above, and from the procedure chosen to estimate the coupling constants in the interaction vertices. In~\cite{jpsi-18} they were obtained by employing  $SU(4)$ symmetry arguments, while here we use the form factors derived from QCD sum rules \cite{nava05}. They introduce the effects of the 
meson off-shellness in the amplitudes and when extrapolated to the  meson on-shell points they yield the coupling constants.  In general, the
couplings involving charm mesons derived from the $SU(4)$  relations agree with the QCDSR estimates within   30 \%  except when pions are 
involved. In this case the discrepancy can be as large as  70 \%. 


In Fig.~\ref{fig5} we also show the time evolution of the ratio  $\mathcal{R}_{\psi}$. The main point here is that we observe a reduction of this ratio during the hadron gas phase. The final ratio becomes roughly  one half of the initial one and this reduction comes mostly from the
reduction of the $\psi(2S)$ abundance also shown in Fig.~\ref{fig5}. 

In  Ref.~\cite{shm-19}  the ratio  $\mathcal{R}_{\psi}$ as a function of the particle transverse momentum  was calculated with the statistical hadronization model. A direct comparison with our results is difficult because the authors of Ref.~\cite{shm-19} did not present the 
$p_T$-integrated yield. They find numbers ranging from $\sim 0.025$, in the low $p_T$ region, to $0.125$, in the high $p_T$ region. The ratio  for low $p_T$ values is close to our result.  In principle these numbers did not have to agree because we include the effects of the hadron gas,
which were neglected in the SHM approach. However, given that we are using the SHM initial conditions for the evolution in the hadron gas, 
and given that this evolution did not change much these numbers, it is reasonable that both predictions are relatively close to each other. 

\begin{figure}[!htbp]
	\centering
\includegraphics[{width=8.0cm}]{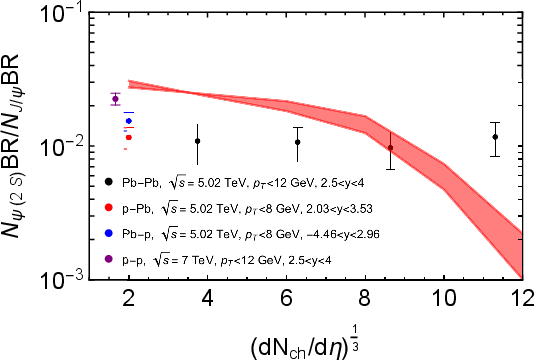} 
\caption{The ratio $(N_{\psi(2S)} BR) / (N_{J/\psi} BR)$ as a function of $\mathcal{N}$. The bands show the uncertainties in the initial
conditions obtained from the SHM. The multiplicities $N_{\psi(2S)}$ and $ N_{J/\psi}$ have been multiplied by the corresponding branching ratios $BR(\psi(2S) \to \mu^+ \mu^- ) = 8 \times 10^{-3}$ and $BR(J/ \psi \to \mu^+ \mu^- ) = 5.961 \times 10^{-2}$~\cite{Workman:2022ynf}. The data have been taken from the ALICE Collaboration for p-p collisions at $7$ TeV~\cite{ALICE:2014cgk} (purple point), for 
p-Pb collisions  at $5.02$ TeV~\cite{ALICE:2014cgk} (blue and red points) and for  Pb-Pb collisions at $5.02$ TeV~\cite {alice24} (black points).  
}
\label{fig6}
\end{figure}

\section{System size dependence}

Now we analyze the dependence of the $\psi(2S)$ and $J/\psi$ final yields and their ratio on a measurable quantity, namely the charged-particle pseudorapidity density at mid-rapidity, $\mathcal{N} \equiv [dN_{ch}/d\eta \,(|\eta|<0.5)]^{1/3}$,  which can be related to the size of the system. As noticed in Ref.~\cite{chiara21},  the freeze-out time $\tau_F$ grows with $\mathcal{N}$. A larger system (either with a larger mass number $A$ and/or in more central collisions and/or at higher energies) produces more particles and larger values of $\mathcal{N}$, which, 
according to the expression found in Ref.~\cite{chiara21}, lasts until a bigger $\tau_{F}$. Hence, the larger the system, the longer is 
the hadron gas phase. 

Next, we will we employ the empirical formulas (power laws) obtained in Ref.~\cite{zczcs} relating 
$\mathcal{N}$ with the volume, charm quark number, charmed meson number and light quark number. Using these relations in (\ref{Eq:Densities}) allows us to estimate the initial conditions for each $\mathcal{N}$ and then solve (\ref{Eq:RatioEquation})  stopping
the evolution at a freeze-out time $\tau_F$ which depends on $\mathcal{N}$. In this way we construct the curve presented in Fig.~\ref{fig6}, showing the  dependence of the ratio  $\mathcal{R}_{\psi}$,  Eq.~(\ref{rat}),  on $\mathcal{N}$. In the figure we compare our 
curve with the experimental data from Refs.~\cite{alice24} and ~\cite{ALICE:2014cgk}.
Since the data from~\cite{alice24} have been given as a function of  $N_{part}$, we have used the relation derived in Ref.~\cite{ALICE:2015juo} to convert them to $\mathcal{N}$. The band shows the uncertainty related to the coupling constants involving $\psi(2S)$.  
We emphasize that  our calculations are performed for the full range in rapidity and transverse momentum of the measured charmonium states, whereas the data refer to restricted $p_T$ and rapidity windows. 
This difference would prevent an appropriate comparison. However, as one can infer from the data, the rapidity dependence of the ratio is 
weak and one may assume that what happens in a part of the phase space also holds for the whole phase space. Under this assumption (and 
knowing that this may be the cause of some discrepancies) we show the comparison between theory and data in Fig.~\ref{fig6}. 

We observe that the ratio decreases with  $\mathcal{N}$. The data come from several recent publications 
~\cite{CMS:2014vjg,Leoncino:2015ooa,CMS:2016wgo,MartinBlanco:2017njc,ALICE:2020vjy,PHENIX:2022nrm,alice24}). 
The only data set which shows a weaker
dependence on the system size is the one fom Ref.~\cite{alice24}. This will certainly deserve further investigation from the theoretical side. 
In our framework this falling behavior of the ratio comes partly from dynamical reasons, since the   
$J/\psi$ and  $\psi(2S)$ interact differently with the hadronic medium. While the first is more regenerated the latter is more absorbed.
The larger the system, the longer the particles live in these conditions and the larger is the $\psi(2S)$ suppression relative to the $J/\psi$. 
However, the dynamical effects are modified by the initial conditions, which control the height of the curve, and by the cooling process, which controls the slope of the curve. The data seem to suggest a smaller initial ratio and also a less steep slope, which would be observed either for a slower (than Bjorken) cooling or for slower varying (with the system size) freeze-out temperature \cite{chiara21}.

Our results can\ be compared to the ratio reported in Ref.~\cite{psi2s}, where we found that $N_{X(3872)}/N_{\psi(2S)}$ grows with the size of the system,  in qualitative agreement with the data from LHCb and CMS Collaborations. If the
geometrical aspect of the interactions were dominant, we would be tempted to relate the ratios   $R_1 = N_{\psi(2S)}/N_{J/\psi}$ and  $R_2 =  N_{X(3872)}/N_{\psi(2S)}$  with the radii of the particles. The fact that $r_{\psi(2S)} > r_{J/\psi}$  implies that $R_1$ decreases with 
$\mathcal{N}$. In a simple molecular model of the  $X(3872)$ we would have $r_{X(3872)} > r_{\psi(2S)}$ implying that $R_2$ would  also 
decrease with $\mathcal{N}$. Since $R_2$ has the opposite behavior, this would suggest that the $X(3872)$ is a compact object, possibly a tetraquark. However, such a
conclusion can only be reached after a careful scrutiny of all the ingredients which lead to the curve shown in 
Fig. ~\ref{fig6}. 

\section{Concluding remarks}
\label{Conclusions}

We have studied the $\psi(2S)$ and $J/\psi$ yield ratio in heavy-ion collisions, taking into account the interactions of the $\psi (2S) $ and $J/\psi$ states with a hadron gas made of light mesons. Starting from effective Lagrangians, we have calculated the thermally-averaged cross sections for the production and absorption of the mentioned states,  which served as input in the rate equations to determine the time evolution of  $N_{\psi(2S)}$, $N_{J/\psi}$ and of $N_{\psi(2S)} / N_{J/\psi}$. The behavior of the ratio is determined mostly by the combined effects of hadronic interactions and hydrodynamical expansion.  Our results agree qualitatively with data.

\begin{acknowledgements}

We are grateful to C. M. Ko, S. H. Lee and R. Rapp for enlightening discussions. 
This work was supported by the Brazilian agencies CNPq, FAPESP and CNPq/FAPERJ (Project INCT-F\'isica Nuclear e Aplica\c{c}\~oes - Contract No. 464898/2014-5). 

\end{acknowledgements}




\end{document}